\documentclass[preprint, aps, nofootinbib]{revtex4}
\usepackage{CJK}
\usepackage{graphicx}%Include figure files
\usepackage{amsmath}
\usepackage{amsfonts}
\usepackage{amssymb}
\usepackage{color}%
\usepackage{dcolumn}% Align table columns on decimal point
\usepackage{slashed}
\usepackage{multirow}
\usepackage{extarrows}
\providecommand{\U}[1]{\protect\rule{.1in}{.1in}}
\newcommand{\f}{\begin{equation}}
\newcommand{\ff}{\end{equation}}
\newcommand{\fa}{\begin{eqnarray}}
\newcommand{\ffa}{\end{eqnarray}}

\begin{document}
\title{Charged Lifshitz black hole and probed Lorentz-violation fermions from holography}
\author{Cheng-Jian Luo $^{1,2}$}
\email{rocengeng@hotmail.com}
\author{Xiao-Mei Kuang$^{3}$}
\email{xmeikuang@gmail.com}
\author{Fu-Wen Shu$^{1,2}$}
\email{shufuwen@ncu.edu.cn}
\affiliation{$^{1}$Department of Physics, Nanchang University, Nanchang, 330031, China\\
$^{2}$Center for Relativistic Astrophysics and High Energy Physics, Nanchang University, Nanchang 330031, China\\
$^{3}$ Instituto de F\'isica, Pontificia Universidad Cat\'olica de Valpara\'iso, Casilla 4059, Valpara\'iso, Chile}

\begin{abstract}
We analytically obtain a new charged Lifshitz solution by adding a non-relativistic Maxwell field in Ho\v{r}ava-Lifshitz gravity. The black hole exhibits an anisotropic scaling between space and time (Lifshitz scaling) in the UV limit, while in the IR limit, the Lorentz
invariance is approximately recovered. We introduce the probed Lorentz-violation fermions into the background and holographically investigate the spectral properties of  the dual fermionic operator. The Lorentz-violation of the fermions will enhance the peak and correspond larger fermi momentum, which compensates the non-relativistic bulk effect of the dynamical exponent ($z$).  For a fixed $z$, when the Lorentz-violation of fermions increases to a critical value, the behavior of the low energy excitation goes from a non-Fermi liquid type to a Fermi liquid type, which implies a kind of phase transition.

\end{abstract}
\maketitle

\section{Introduction}
The Anti-de Sitter/Conformal Field Theory (AdS/CFT) correspondence\cite{maldacena1,agmoo,gkp,witten} opens a new avenue to disclose many different strongly interacting systems. The correspondence makes a connection between  gravitational theories on AdS spacetime and a quantum field theory(QFT) that lives on the conformal boundary of the AdS spacetime. Due to its remarkable feature connecting a strong coupling QFT with a weak coupling gravitational theory, this useful tool has been attracting considerable interest in studying strongly coupled physics, especially, the possible applications to condensed matter physics(see \cite{hartnoll,herzog,mcgreevy,Zaanen} for reviews).

The AdS/CFT duality is first generated in \cite{Kachru:2008yh} to study a boundary theories with  dynamical critical exponents which show the dynamical scaling
\begin{equation}\label{eq-lifshitzScale}
t\to \lambda^z t,~x\to \lambda x~~\mathrm{with}~~ z\neq 1
\end{equation}
instead of the scale invariance with $z=1$.  It is pointed out that the background, which is holographically applied to study the nonrelativistic  QFT with dynamical (Lifshitz) scaling \eqref{eq-lifshitzScale}, is given by the Lifshitz spacetime
\begin{eqnarray}\label{eq-LifshitzMetric}
ds^2=-r^{2z}dt^2+\frac{dr^2}{r^2}+r^2dx_i^2,
\end{eqnarray}
where the spacial index $i$ runs from 1 to $D-2$ and here we set the radius of curvature to be unit.

In \cite{Kachru:2008yh}, though the dual QFT is non-relativistic with Lifshitz scaling, the bulk gravity is relativistic which are invariant under the full spacetime diffeomorphism and it has to couple with  matters to generalize the Lifshitz geometry.  This situation has been improved by the Ho\v{r}ava-Lifshitz(HL) gravity \cite{Horava:2009uw,Horava:2011gd} which is based on the perspective that Lorentz symmetry should appear as an emergent symmetry at long distances, but can be fundamentally absent at short distances \cite{TP,CN}. HL theory is a non-relativistic power-counting renormalizable theory of gravitation.
Since the HL gravity is itself anisotropic between space and time, it is natural to expect that the HL gravity provides a minimal holographic dual for non-relativistic Lifshitz-type QFT. This proposal has been realized and carefully studied in \cite{Griffin:2012qx,Janiszewski:2012nf,Janiszewski:2012nb}.  Especially, it was addressed in \cite{Griffin:2012qx} that the Lifshitz spacetime is a vacuum solution to the HL gravity.  Analytical vacuum solution with asymptotically Lifshitz geometry was also found in \cite{Alishahiha:2012iy,wang2014}. The holographic applications in the framework of the HL gravity were constructed (for instance, holographic superconductor models in the HL gravity were studied in \cite{Lin,Lin1,Luo} ).

In this paper, as a first step, we couple the gravitational action of HL gravity with a $U(1)$ gauge field whose Lagrangian is consistent with the  symmetry of HL gravity. We analytically solve the equations of motions for the gravity and the gauge field, and obtain a new charged Lifshitz black hole solution to the coupled theory. As we will show later, in the UV limit, our charged solution behaves as the geometry \eqref{eq-LifshitzMetric}  while at the IR limit, the Lorentz invariance (with $z = 1$) is restored and the geometry is $AdS_2\times\mathbb{R}^{D-2}$.

Then we introduce the Lorentz-violation fermions in the charged HL background and  study the properties of fermionic spectral. The motivations we introduce the Lorentz-violation Dirac field  stem from the following two aspects. On one hand,  the authors of \cite{Lopes:2015bra} proposed that the general low energy action of the HL type can be derived as the spectral action for Dirac operator with P-violating terms. They also addressed that the obtained general Lorentz violating fermionic action related with the Dirac operator is consistent with the fermionic action of Standard Model Extension (SME) studied in \cite{SMEaction},  and the Lorentz violating effects is very small in weak gravitational field. Besides, in HL theory, the Lorentz symmetry is kind of approximate symmetry at low energy, which motivates us to consider Dirac field with weak Lorentz violation in this background.
On the other hand, holography has been widely applied  to explore the strongly correlated fermionic system.  In the pioneer works \cite{f1,f2,f3,f4,IL} on this topic, the authors proposed that by studying the bulk Dirac equation in the RN-AdS black hole with the ingoing behavior at the horizon, the fermionic correlation can be extracted at the AdS boundary. The proposal helps us further understand  the mysterious behaviors of the existing (non-)Fermi liquid. Besides in relativistic UV theory,  the fermionic response was also studied in Lifshitz theory in \cite{Hartnoll:2011dm,Cubrovic:2011xm}. These works has inspired  more and more efforts to be involved in the related topics\footnote{The generalization of holographic fermions obtain many
remarkable progress, such as the effects of different bulk theory on the fermionic correlation\cite{JPW1,kuang1,kuang2,JPW2,1201.1764,FLQ1,FLQ2,FLQ3,Fan:2013tpa,Fan:2013zqa}, holographic non-relativistic fermionic fixed points by imposing the Lorentz violating boundary condition\cite{1108.1381,1110.4559,1111.3783,Lin2,1409.2945},  phase transition from the Fermi liquid to non-Fermi liquid and also to the Mott insulating phase due to the interaction between Dirac field and gauge field\cite{1010.3238,1012.3751,1102.3908,1405.1041,1404.4010,1411.5627} and the fermionic spectral function effected by lattice effects\cite{Hartnoll:2012rj,Liu:2012tr,1304.2128,1410.7323,Fang:2015dia}, etc..}.
However, the previous works usually started by adding relativistic fermions into the (non-)relativistic bulk theory. Since the anisotropic scaling \eqref{eq-lifshitzScale} is built by construction in the HL gravity, it is natural to expect the coupling matters exhibit the same scaling in the same footing. This is the other motivation to introduce the probed non-relativistic fermions into the non-relativistic HL gravity and study the effect of the Lorentz-violation on the Fermionic spectral function.

Our results show that,  different from the  effect of dynamical exponent which suppresses the peak of the fermionic spectral function\cite{FLQ1,1409.2945}, the Lorentz-violation of fermions will  enhance the peak and make the fermi momentum larger to compensate the non-relativistic bulk effect. Also, for fixed $z$, small Lorentz-violation of fermions corresponds the low energy excitation as non-Fermi liquid with nonlinear dispersion relation, when the Lorentz-violation becomes larger than a critical value, the low energy excitation always behaves as Fermi liquid with linear dispersion relation. This may imply that in the non-relativistic HL gravity, it is more natural to introduce Lorentz-violation of fermions to dually describe a fermionic system whose low energy excitation is Fermi liquid.

The remaining of this paper is organized as follow. In section \ref{sec-solution}, we obtain a charged black hole solution by solving the
coupled system of non-relativistic gauge field and HL gravitational theory.  We show the holographic setup by introducing the Lorentz-violation fermions into the coupled background sector in section \ref{sec-setup}.  Then in section \ref{sec-result}, we exhibit  the numerical results of the Fermi momentum and dispersion relation by analyzing the spectral function. Section \ref{sec-conclusion} contributes to our conclusion and discussion.

\section{Analytical charged Lifshitz solution from Horava-Lifshitz Gravity}\label{sec-solution}
It was studied in Ref. \cite{Griffin:2012qx} that the HL gravity can provide solution dual to Lifshitz-type field theories, where the following gravitational action is proposed
\begin{equation}\label{eq-Gaction}
S_{g}=\frac{1}{2\kappa^2}\int dt dr d^dx\sqrt{-G}(K_{ab}K^{ab}-\lambda K^2+\beta(R-2\Lambda)+\frac{\alpha^2}{2}\frac{\nabla_aN\nabla^aN}{N^2}).
\end{equation}
It will return to Einstein gravity as $\lambda=\beta=1$ and $\alpha=0$. In Ref. \cite{Griffin:2012qx}, the authors found that for the requirement of stability and perturbative unitarity around flat spacetime, constraints should be imposed on the couplings, i.e., $\beta>0$, $\alpha\leq\frac{2\beta d}{d-1}$ and $\lambda\geq1$ or $\lambda^2\leq\frac{1}{d-1}$ for $\Lambda=0$. When $\Lambda<0$, if we take
\begin{eqnarray}\label{eq-Lambda}
\Lambda=-\frac{(d+z-1)(d+z)}{2},~and~\alpha^2=\frac{2\beta(z-1)}{z},
\end{eqnarray}
the Lifshitz metric (\ref{eq-LifshitzMetric}) is also a vacuum solution of the equation (\ref{eq-Gaction}). In our paper, we will use the condition with $\Lambda<0$. Besides, the authors worked with the Arnowitt-Deser-Misner metric ansatz,
\begin{equation}\label{eq-metric1}
ds^{2}=-N^{2}dt^{2}+g_{ij}(dx^{i}-N^{i}dt)(dx^{j}-N^{j}dt).
\end{equation}
Here, $N$, $N^{i}$ and $g_{ij}$ are the lapse function, the shift vector and the metric of the space-like hypersurface, respectively. Thus, in the action, one has the formula $\sqrt{-G}=\sqrt{g}N$,  $K_{ab}=\frac{1}{2N}(\partial_tg_{ab}-\nabla_aN_b-\nabla_bN_a)$ and $K=g^{ab}K_{ab}$, where $R$ is the scalar curvature of the metric $g_{ab}$.
Later, the authors of \cite{Alishahiha:2012iy} analytically studied the neutral Lifshitz black hole solutions to the gravitational thoery.

Here we intend to explore the possible charged black hole solutions in HL gravity. We consider Lorentz-violating electromagnetism field coupling to the HL gravity with the action \cite{Kimpton:2013zb}
\begin{eqnarray}\label{eq-Maction}
\nonumber S_{m}=-\frac{1}{2\kappa^2}\int dt dr d^dx\sqrt{g}N && \big(\frac{2}{N^{2}}g^{ij}(F_{0i}-F_{ki}N^{k})(F_{0j}-F_{\ell j}N^{\ell})-F_{ij}F^{ij}\\
&&-\beta_{0}-\beta_{1}a_{i}B^{i}-\beta_{2}B_{i}B^{i}\big),
\end{eqnarray}
where $F_{\mu\nu}=\partial_\mu A_\nu-\partial_\nu A_\mu$ and $B^i=\frac{1}{2}\frac{\epsilon^{ijk}}{\sqrt{g}}F_{jk}$ with $\epsilon^{ijk}$ the Levi-Civita symbol. Then the total action of the background we will take account into is
\begin{equation}\label{eq-Taction}
S_t=S_g+S_m.
\end{equation}
We consider the electromagnetic field with the only non-vanishing component $A_t(r)$ and also $\beta_\mu=0(\mu=0,1,2)$. Subsequently, the Maxwell equation reads as
\begin{eqnarray}\label{eq-gaugeF}
\partial_r(\sqrt{g}NF^{rt})=0,
\end{eqnarray}
which gives us the solution
\begin{eqnarray}
F^{rt}=\frac{Q_e}{\sqrt{g}N}
\end{eqnarray}
with the integral constant $Q_e$, which can be interpreted as the charge of the Lifshitz black hole.

In order to get the complete solution to the equations of motion deduced from the action (\ref{eq-Taction}), we will borrow the analytical process shown in  Ref. \cite{Alishahiha:2012iy}. To proceed, we set the metric components in (\ref{eq-metric1}) as
\begin{eqnarray}\label{eq-metricproduct}
N=e^{2f(r)},~g_{rr}=\frac{1}{e^{2h(r)}},~g_{ii}=e^{2l(r)},~N_a=0.
\end{eqnarray}
By substituting the formulas in  (\ref{eq-metricproduct}) into the action (\ref{eq-Taction}), we obtain
\begin{eqnarray}
S_t=&&\frac{d\beta\upsilon}{2\kappa^2}\int dr e^{dl+h+f}\left((d-1)l'^2+2l'f'+\delta f'^2-\frac{2\Lambda}{d}e^{-2h}-\frac{2Q_e^2}{d\beta}e^{-2h-2dl}\right),
\end{eqnarray}
where $\upsilon=\int dtd^dx$ and
\begin{eqnarray}\label{eq-delta}
\delta=\frac{\alpha^2}{2d\beta}.
\end{eqnarray}
Considering the above $S_t$ as  one dimensional action of the
functions $f(r)$, $l(r)$ and $h(r)$ and then variating the action, we can get two independent equations
\begin{eqnarray}\label{eq-motionoflf2}
&&\left(e^{dl+h+f}(l'+\delta f')\right)'=-\frac{2Q_e^2}{d\beta}e^{f-h-dl}-\frac{2\Lambda}{d}e^{f-h+dl},\\
&&\left(e^{dl+h+f}((d-1)l'+f')\right)'=-2\Lambda e^{f-h+dl},
\end{eqnarray}
which can be further simplified as
\begin{eqnarray}\label{eq-motionoflf3}
\left(e^{dl+h+f}((d\delta-1)f'+l')\right)'=-\frac{2Q_e^2}{\beta}e^{f-h-dl}.
\end{eqnarray}
It seems not so direct to find the solutions of the three functions $f(r)$, $l(r)$ and $h(r)$, but we can assume the following forms
\begin{eqnarray}\label{eq-lfhansatz}
f=z\ln r+\frac{1}{2}\ln \xi(r),~h=\ln r+\frac{1}{2}\ln \xi(r),~l=\ln r,
\end{eqnarray}
so that the metric becomes
\begin{eqnarray}\label{eq-metric0}
ds^2=-r^{2z}\xi(r)dt^2+\frac{dr^2}{r^2\xi(r)}+r^2dx_i^2.
\end{eqnarray}
The issue now reduces to find the solution of $\xi(r)$. Inserting the formulas (\ref{eq-lfhansatz}) into equation (\ref{eq-motionoflf3}), gives us the equation of motion for $\xi(r)$
\begin{eqnarray}
\left(\frac{1}{2z}r^{d+z+1}\xi'(r)\right)'=-\frac{2Q_e^2}{\beta}r^{z-1-d}.
\end{eqnarray}
It is obvious that  if we take the constraints (\ref{eq-Lambda}), then we will get the solution of $\xi(r)$
\begin{eqnarray}\label{eq-metric}
\xi(r)=1-\frac{M}{r^{d+z}}+\frac{2zQ_e^2}{d\beta(d-z)}\frac{1}{r^{2d}},
\end{eqnarray}
where $M$ and $Q_e$ are integral constants which can be understood as the mass and charge of the Lifshitz black hole. At the horizon $r_0$,
$\xi(r_0)$ should be zero, which implies that $M=r_0^{d+z}+\frac{2zQ_e^2}{d\beta(d-z)}\frac{1}{r_0^{d-z}}$.  Substituting the solved metric into (\ref{eq-gaugeF}), we can fix the Maxwell field as
\begin{eqnarray}\label{eq-gaugeA}
A_t=\mu\left(1-(\frac{r_0}{r})^{d-z}\right)~~\text{with}~~\mu=\frac{Q_e}{d-z}.
\end{eqnarray}
The Hawking temperature of the charged black brane is
\begin{eqnarray}\label{eq-T}
T=\frac{(d+z)r_0^z-\frac{2z}{d\beta}Q_e^2r_0^{z-2d}}{4\pi}.
\end{eqnarray}

We will move on to give some comments on our black hole solution.

Firstly, the solution (\ref{eq-metric0}), (\ref{eq-metric}) and (\ref{eq-gaugeA}) is a new charged Lifshitz solution which is different from the solutions proposed in\cite{Taylor:2008tg,dugui2,dugui3,dugui4,dugui5,dugui6,1401.6479}. However, it  can recover AdS RN black hole solution when $z=1$.  It is worthy to point out that from the expression of $\delta$ in equation (\ref{eq-delta}), $\alpha=0$ will force us to keep $z=1$ which is the AdS RN solution. Thus we can only expect the above Lifshitz solution when $\alpha\neq 0$, which also happens in neutral case studied in \cite{Alishahiha:2012iy}.

Secondly, from the expression of the temperature (\ref{eq-T}), it is straightforward that the extremal case is fulfilled when
\begin{eqnarray}\label{eq-Q-mu}
Q_e=\sqrt{\frac{d\beta(d+z)}{2z}}r_0^d,~~ \text{or} ~~\mu=\frac{1}{(d-z)}\sqrt{\frac{d\beta(d+z)}{2z}}r_0^d.
\end{eqnarray}
Then it is explicit that in order to have real chemical potential, the condition $\beta>0$ should be satisfied. Furthermore, recalling the expression of $\delta$ in equation (\ref{eq-delta}), we should have the range of the Lifshitz exponent as $z>1$ with $\alpha\neq 0$.

Thirdly,  we will check the asymptotical behavior of our solution. In asymptotical boundary $r\to\infty$, since $\xi(r\to\infty)\to1$, so that the metric (\ref{eq-metric0}) reduces to the asymptotical Lifshitz metric \eqref{eq-LifshitzMetric}.  While, near the horizon, the geometry is a bit subtle. For simplicity, we will set the horizon $r_0=1$ in the following discussion, then the redshift factor at zero temperature
can be rewritten as
\begin{eqnarray}\label{eq-metric2}
\xi(r)=1-\frac{2d}{(d-z)r^{d+z}}+\frac{d+z}{(d-z)r^{2d}}.
\end{eqnarray}
whose behavior near the horizon is $\xi(r)\simeq2d(d+z)(r-1)^2=\frac{1}{L_2^2}(r-1)^2$ with $L_2=1/\sqrt{2d(d+z)}$. Therefore, at the zero temperature, the near horizon geometry is $AdS_2\times\mathbb{R}^d$ with the AdS radius $L_2$. This is more explicit by considering the transformation
\begin{eqnarray}
r-1=\lambda\frac{L_2^2}{\varsigma},~t=\lambda^{-1}\tau,~\lambda\rightarrow 0~with~\varsigma,~\tau~finite.
\end{eqnarray}
Consequently, the metric (\ref{eq-metric0}) can be deduced into the $AdS_2\times\mathbb{R}^d$ form
\begin{eqnarray}
ds^2=\frac{L_2^2}{\varsigma^2}(-d\tau^2+d\varsigma^2)+dx_i^2
\end{eqnarray}
and the gauge field is
\begin{eqnarray}
A_\tau=\frac{(d-z)\mu L_2^2}{\varsigma}.
\end{eqnarray}

Since the near horizon geometry is $AdS_2\times\mathbb{R}^d$ and the solution is Lifshitz in UV limit, we can use the Lifshitz holography\cite{Kachru:2008yh} to study its application in strongly coupled physics. In the next section, as an attempt, we will study the fermionic spectral of the operator dual to the probed fermions in this background .

%%%%%%%%%%%%%%%%%%
\section{Holographic setup of Lorentz violation fermions}\label{sec-setup}
We will introduce the fermions into the above theory as a probe and apply the holography to study the fermionic correlation. We consider the Lorentz-violation action for the probed fermionic sector \cite{SMEaction}
\begin{eqnarray}\label{eq-smeation}
S=\int d^{d+2}x\sqrt{g}\bar{\psi}(\Gamma^{a}e_{a}^{~\mu}(\nabla_{\mu}-iqA_{\mu})+M)\psi,
\end{eqnarray}
with
\begin{eqnarray}
\Gamma^{a}&=&\gamma^{a}-c_{\mu\nu}e^{a\nu}e_{b}^{~\mu}\gamma^{b}-d_{\mu\nu}e^{a\nu}e_{b}^{~\mu}\gamma^5\gamma^b-e_{\mu}e^{a\mu}-if_{\mu}e^{a\mu}
-\frac{1}{2}g_{\lambda\mu\nu}e^{a\nu}e_{b}^{~\lambda}e_{c}^{~\mu}\gamma^{bc}\nonumber,\\
M&=&m_0+im_5\gamma^5+m_{\mu}e_{a}^{~\mu}\gamma^{a}+b_{\mu}e_{a}^{~\mu}\gamma^5\gamma^a+\frac{1}{2}H_{\mu\nu} e_{a}^{~\mu}e_{b}^{~\nu}\gamma^{ab}\nonumber,\\
\nabla_{\mu}&=&\partial_{\mu}-\frac{1}{4}\omega_{\mu ab}\gamma^{ab},~~
\gamma^{ab}=\frac{1}{2}[\gamma^a,\gamma^b],
\end{eqnarray}
The first term in $\Gamma^{a}$ and the first two terms in $M$ respectively lead to the usual Lorentz- invariant kinetic term and  mass term for the Dirac field. In order to take Lorentz-violation $\Gamma^a$ and $M$, we will choose the setting of parameters as $c_{\mu\nu}=\rho h_{\mu\nu}$ where $h_{\mu\nu}=g_{\mu\nu}+n_{\mu}n_{\nu}$ and $n_{\mu}=(-N,0,0,0)$, $H_{tr}=H_{rt}=i\eta$, $m_0=-m=const.$, and the other coefficients vanish.

From the action (\ref{eq-smeation}), the equation of motion for Dirac field is
\begin{eqnarray}\label{eq-Dirac}
(\Gamma^{a}e_{a}^{~\mu}(\nabla_{\mu}-iqA_{\mu})+M)\psi=0.
\end{eqnarray}
Redefining the field as
\begin{eqnarray}
\psi=(-gg^{rr})^{\frac{1}{4}}\int d\omega dke^{-i\omega t+ik_{i}x^{i}}\phi,
\end{eqnarray}
and setting $\phi=(\phi_{1},\phi_{2})^{T}$ with $\phi_I$ the two-component spinors, then choosing $k_{i}=k \delta_x^{i}$ because of the rotation symmetry in the spatial directions, we reduce the Dirac equation (\ref{eq-Dirac}) into the form
\begin{eqnarray}\label{eq-diracofphi}
(\frac{\sqrt{g_{xx}}}{\sqrt{g_{rr}}}\partial_r+m\sigma^3\sqrt{g_{xx}})\phi_I=(\frac{\sqrt{g_{xx}}}{\sqrt{g_{tt}}}(\omega+qA_t)i\sigma^2
\mp k\sigma^1-\frac{\eta}{2a}\sigma^1\sqrt{\frac{ g_{xx}}{g_{tt}g_{rr}}})\phi_I,
\end{eqnarray}
with $a=(1-(d+1)\rho)$ after we choose the gamma matrices as \cite{gammamatrices}
\begin{eqnarray}\label{eq-Gamma}
\Gamma^{r}=\left(
\begin{array}{cc}
-\sigma^3 & 0\\
0 & -\sigma^3
\end{array}
\right),~
\Gamma^{t}=\left(
\begin{array}{cc}
i\sigma^1 & 0\\
0 & i\sigma^1
\end{array}
\right),~
\Gamma^{x}=\left(
\begin{array}{cc}
-\sigma^2 & 0\\
0 & \sigma^2
\end{array}
\right).
\end{eqnarray}

Further setting $\phi_I=(X_{I},Y_{I})^T$, we  divide  the above equation (\ref{eq-diracofphi}) into two equations
\begin{eqnarray}
\frac{\sqrt{g_{xx}}}{\sqrt{g_{rr}}}\partial_rX_I+m\sqrt{g_{xx}}X_I&=&\frac{\sqrt{g_{xx}}}{\sqrt{g_{tt}}}(\omega+qA_t)Y_I
\mp kY_I-\eta\sqrt{\frac{g_{xx}}{g_{tt}g_{rr}}}Y_I, \\
\frac{\sqrt{g_{xx}}}{\sqrt{g_{rr}}}\partial_rY_I-m\sqrt{g_{xx}}Y_I&=&-\frac{\sqrt{g_{xx}}}{\sqrt{g_{tt}}}(\omega+qA_t)X_I
\mp kX_I-\eta\sqrt{\frac{g_{xx}}{g_{tt}g_{rr}}}X_I.
\end{eqnarray}
Not that we have shifted  $\frac{\eta}{2a}\to \eta$ in the equations. Later, we will use $\eta$ to denote the strength of the Lorentz-violation of the fermions, and $\eta=0$ means that the fermions are Lorentz-invariance. With the above equations in hands, we further introduce the new ratios $\chi_I=\frac{X_I}{Y_I}$. Hereafter, the Dirac equation can be reduced into
\begin{eqnarray}\label{eq-diracofki}
(r^2\sqrt{\xi(r)}\partial_r+2mr)\chi_I&=&\frac{1}{r^{z-1}\sqrt{\xi(r)}}(\omega+qA_t)+ (-1)^Ik-\frac{\eta}{r^{z-2}}\nonumber\\
&+&\left(\frac{1}{r^{z-1}\sqrt{\xi(r)}}(\omega+qA_t)- (-1)^Ik+\frac{\eta}{r^{z-2}}\right)\chi_{I}^2.
\end{eqnarray}
The boundary condition of $\chi_I$ near the horizon $r\rightarrow 1$ is
\begin{equation}
\chi=i.
\end{equation}

{\bf UV behavior}

In UV limit, the metric (\ref{eq-metric0}) becomes
\begin{equation}
g_{tt}\rightarrow- r^{2z},~g_{rr}\rightarrow\frac{1}{r^2},~g_{xx}=r^2,
\end{equation}
with $z>1$ and the equation (\ref{eq-diracofphi}) becomes
\begin{equation}\label{eq-UV}
(r^2\partial_r+mr\sigma^3-r^{1-z}(\omega+qA_t)i\sigma^2-(-1)^Ik\sigma^1-\eta r^{2-z}\sigma^1)\phi_I=0.
\end{equation}
of which the leading order\footnote{When $z\rightarrow1$, the leading order is too stiff to say (\ref{eq-leadings}) is the leading order of equation (\ref{eq-UV}). In fact, later in our numerical calculation, we always choose $z=1+\epsilon$ where $\epsilon$ is not a small number, it cannot be ignored. So when $r\rightarrow\infty$, it will be general to choose the leading order like equation (\ref{eq-leadings}).} is
\begin{eqnarray}\label{eq-leadings}
(\partial_r+\frac{m\sigma^3}{r})\phi_I=0,
%,~z\geq1+\epsilon;\\
%(\partial_r+\frac{m\sigma^3-\eta\sigma^1}{r})\phi_I=0,~z=1;
\end{eqnarray}
%where $\epsilon$ is a suitable number which will let $\frac{\eta\sigma^1}{r^{z}}(z>1+\epsilon)$ negligible. With the above equations, we can find that when $\eta\neq0$ and $z=1+\epsilon$, the parameters of $\phi_1$ and $\phi_2$ will associate with each other. As discussed above, in our paper, we will take more attention at $z\geq1+\epsilon$ to take the relationship between the Green function and the parameter $\eta$ and the dynamic critical exponent $z$. So when $z\geq1+\epsilon$,
and its solution near the boundary is
\begin{eqnarray}\label{eq-SR}
\phi_I=a_Ir^{-m}\left(
\begin{array}{c}
1\\
0
\end{array}
\right)+b_Ir^{m}\left(
\begin{array}{c}
0\\
1
\end{array}
\right)~~\text{with}~r\rightarrow\infty.
\end{eqnarray}
As it is addressed in \cite{IL}, choosing the normal boundary term
\begin{eqnarray} \label{boundary2}
S_{bdy}&=&\frac{i}{2}\int_{\partial\mathcal{M}}dtd^dx\sqrt{-gg^{rr}}\bar{\phi}\phi~,
\end{eqnarray}
we can suppose that the source and the response in (\ref{eq-SR}) are related by
\begin{eqnarray}\label{eq-SR1}
a_I\left(
\begin{array}{c}
1\\
0
\end{array}
\right)=Sb_I\left(
\begin{array}{c}
0\\
1
\end{array}
\right),
\end{eqnarray}
and the boundary Green's function $G(\omega,k)$ is
\begin{eqnarray}
G=-iS\gamma^0.
\end{eqnarray}
Recalling the definition of $\chi_I=\frac{X_I}{Y_I}$, we can rewrite the Green function as
\begin{eqnarray}\label{eq-GF}
G(\omega,k)=\left(
\begin{array}{cc}
G_1&0\\
0&G_2
\end{array}
\right)
=\lim_{r\rightarrow\infty}r^{2m}\left(
\begin{array}{cc}
\chi_1&0\\
0&\chi_2
\end{array}
\right).
\end{eqnarray}

{\bf IR behavior}

In near-horizon geometry $AdS_2\times\mathbb{R}^d$, the Dirac equation (\ref{eq-diracofphi})  in
the zero frequency limit or low energy limit, can be rewritten as
\begin{eqnarray}
\varsigma\partial_\varsigma\phi_I-(mL_2\sigma^3-i(d-z)q\mu L_2^2\sigma^2-(-1)^IL_2k\sigma^1-
\frac{\eta\varsigma^2\sigma^1}{L_2})\phi_I=0,
\end{eqnarray}
where $\Gamma^\varsigma=-\Gamma^r$ and others are the same as (\ref{eq-Gamma}) when we reflect the change between the radial coordinate $r$ to the coordinate $\varsigma$. Moreover, near the $AdS_2$ boundary with $\varsigma\rightarrow0$, the leading term of the above equation is
\begin{eqnarray}\label{eq-eqads2B}
\varsigma\partial_\varsigma\phi_I=\left(
\begin{array}{cc}
mL_2&-(d-z)q\mu L_2^2-(-1)^IL_2k\\
(d-z)q\mu L_2^2-(-1)^IL_2k&-mL_2
\end{array}
\right)\phi_I,
\end{eqnarray}
which gives us the leading behaviour of $\phi_I$ as
\begin{eqnarray}
\phi_I=b_I^{(0)}\upsilon_-\varsigma^{-\nu(k)}+a_I^{(0)}\upsilon_+\varsigma^{\nu(k)},
\end{eqnarray}
where $\upsilon_\pm$  and $\pm\nu(k)$ are real eigenvectors and eigenvalues of the matrix in the right side of  equation \eqref{eq-eqads2B} with the form
\begin{eqnarray}\label{eq-nu}
\nu(k)=\sqrt{(m^2+k^2)L_2^2-(d-z)^2q^2\mu^2 L_2^4}.
\end{eqnarray}

Subsequently, according to the discussion in \cite{f3}, in the IR CFT
the conformal dimension of the operator $\mathbf{O}_k$ will be  $\delta_k = 1/2 + \nu(k)$, which is dependent of
 the boundary dimension and Lifshitz dynamical critical exponent because they have information in the IR geometry.

%%%%%%%%%%%%%%%%%%
\section{Results of fermi surface and dispersion relation}\label{sec-result}
In this section, we will numerically solve the Dirac equation \eqref{eq-diracofki} to investigate the momentum of Fermi surface and the dispersion relation of the excitation near the Fermi surface. In the previous works\cite{FLQ1,1409.2945}, the authors claimed that as the Lifshitz exponent $z$ increases, the peak of the Fermi-like peak suppresses and for large enough $z$, the peak is too small to be observed. In our model, the same phenomena is observed. Thus, we will fix a small $z=1.02$\footnote{Since $z$ suppresses the peak of Green function, we do not choose it more different from 1, which would be more interesting because the system is more  non-relativistic.} and mainly focus on  the effects of the Lorentz violation parameter $\eta$ on the holographic fermionic
systems. Meanwhile, we will fix the mass of fermions $m=0$ and charge $q=1$ in the $3+1(d=2)$ dimensional bulk geometry\footnote{Note that effect of dimension of theory and fermionic mass as well as charge on holographic fermions has been carefully investigated in some gravitational theories\cite{kuang1,1409.2945}.}.

Then from equation \eqref{eq-diracofki}, the Green function satisfies the following symmetry
\begin{equation}\label{eq-symmetry}
G_1(\omega,k)=G_2(\omega,-k),~~~~~G_I(\omega,k)=-1/G_I(\omega,-k),
\end{equation}
so that without loss of generality, we can only study the component $G_2$ with positive $k$ for the Fermi surface and dispersion relation. Note that we will study at zero temperature with $\mu$ having the form in equation \eqref{eq-Q-mu}.

\begin{figure}[h]
  \centering
  \includegraphics[width=.3\textwidth]{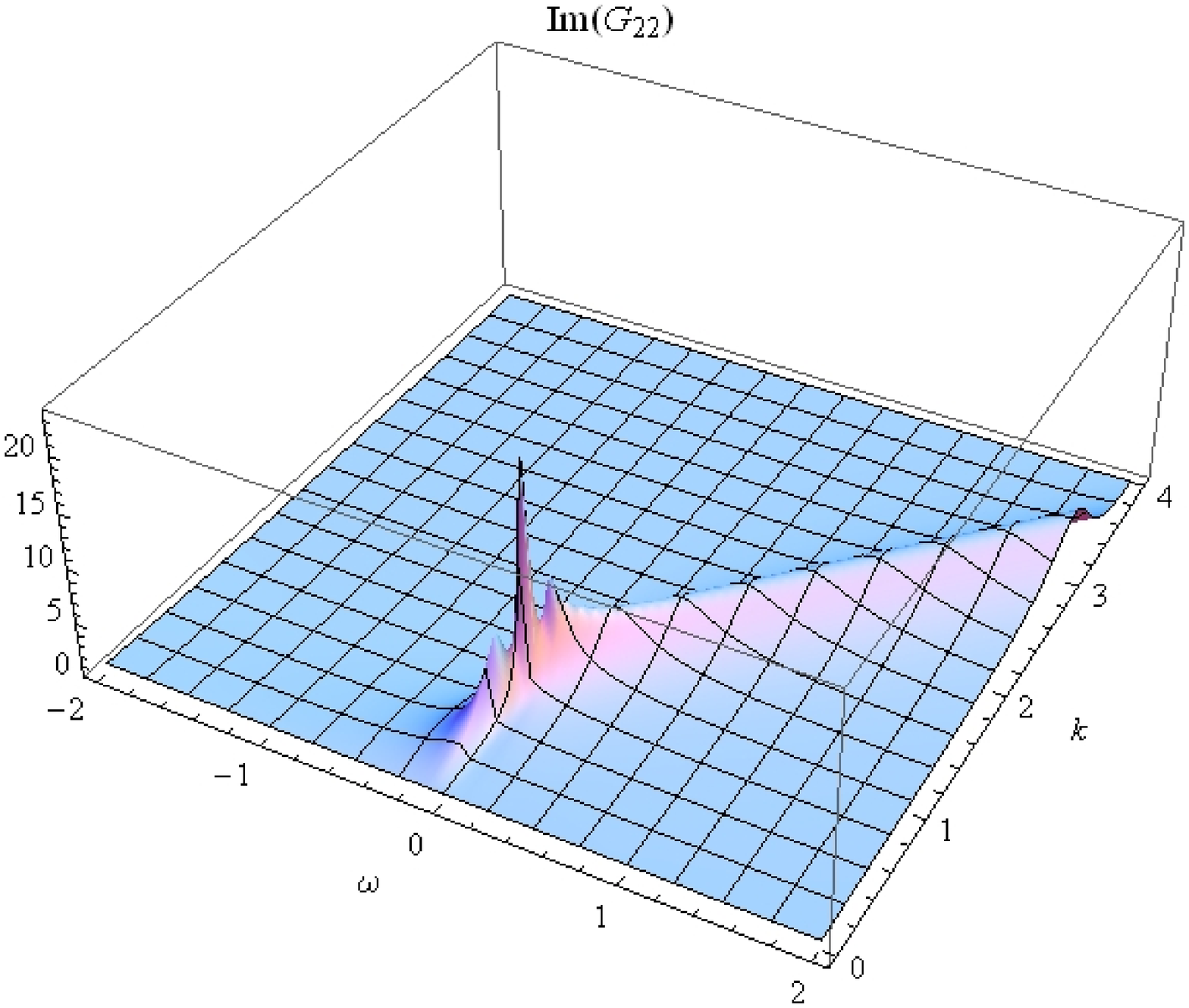}\hspace{0.3cm}
  \includegraphics[width=.3\textwidth]{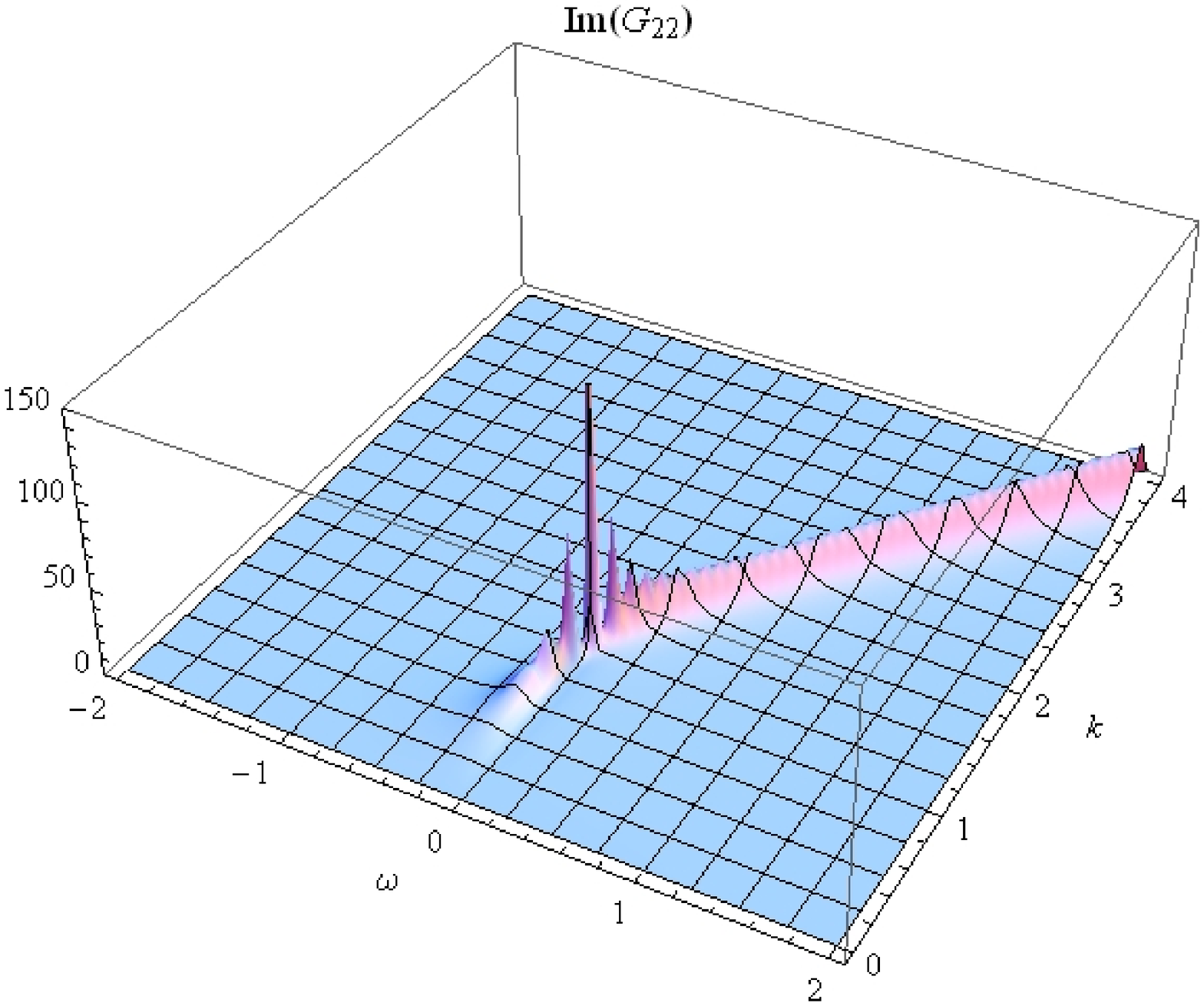}\hspace{0.3cm}
  \includegraphics[width=.3\textwidth]{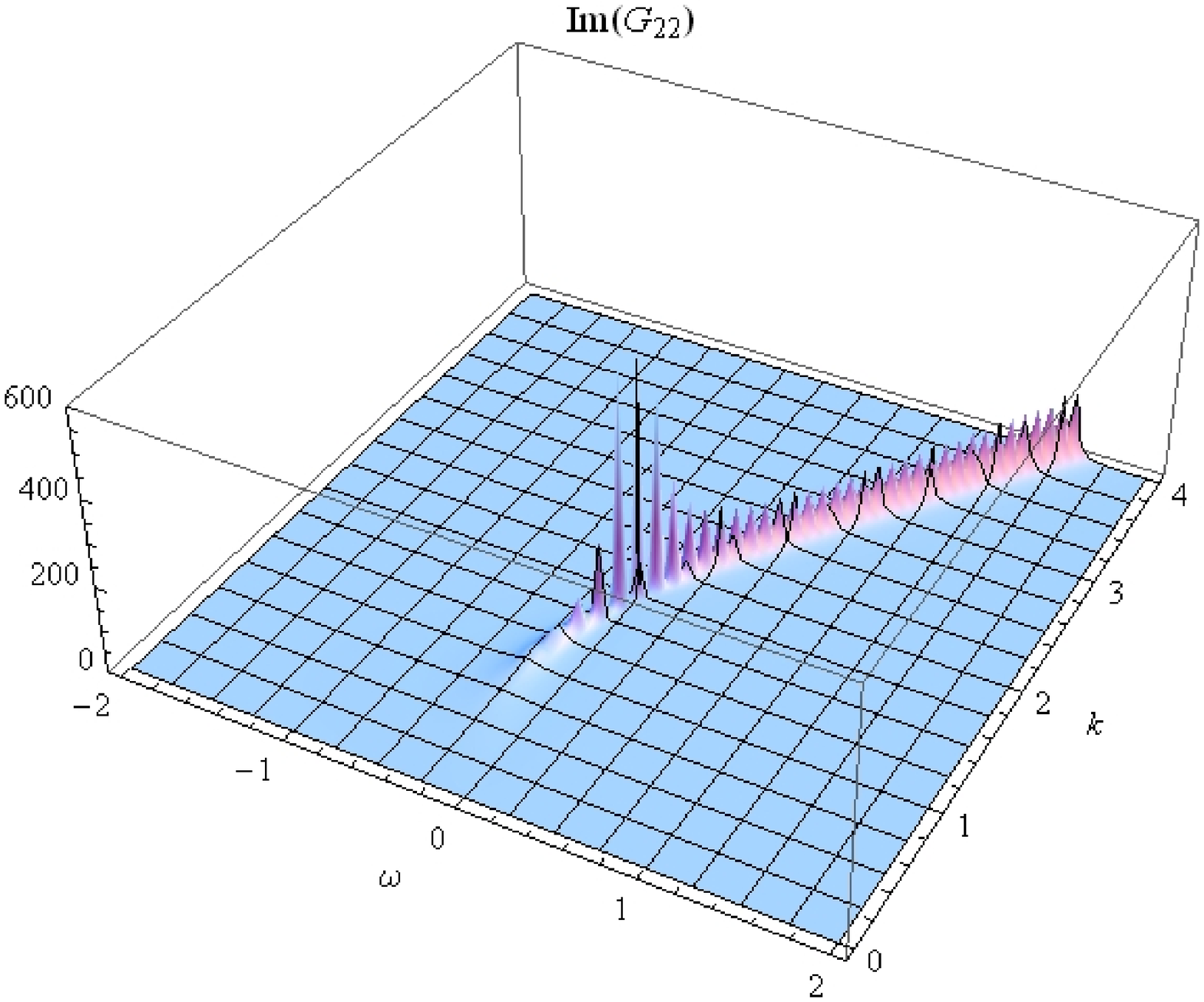}\hspace{0.3cm}
  \caption{The Green function with $\eta=0$(left), $\eta=0.03$(middle), $\eta=0.05$(right).}\label{fig-3D}
\end{figure}
\begin{figure}[h]
  \centering
  \includegraphics[width=.4\textwidth]{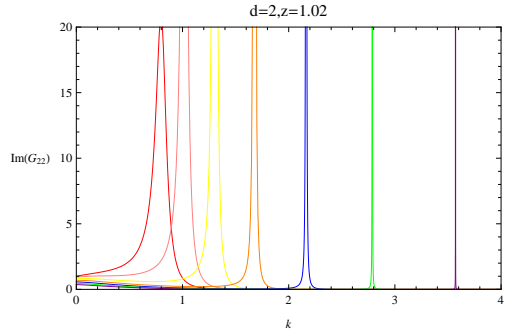}\hspace{1cm}
  \caption{Green function changing as the momentum with tiny $\omega=10^{-5}$ for different $\eta$. From left to right, we increase $\eta$ from $0$ to $0.06$ with the step $0.01$.}\label{fig-d2z15etac}
\end{figure}

\begin{figure}[h]
  \centering
  \includegraphics[width=.4\textwidth]{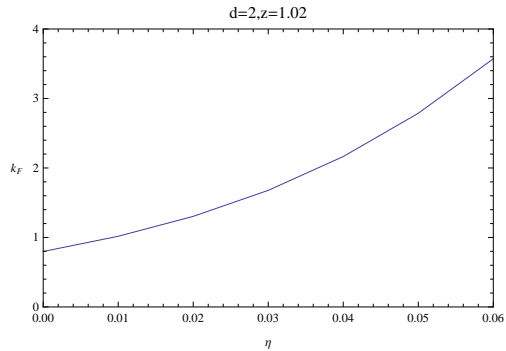}\hspace{1cm}
  \caption{The change of Fermi momentum $k_F$ as $\eta$.}\label{fig-d2z15etakf}
\end{figure}

In figure \ref{fig-3D}, we show how the Green function $G_{2}$ behaves as the momentum and frequency for different $\eta$. The left plot corresponds to $z=1.02$ and $\eta=0$, which shows a sharp Fermi-like peak at zero frequency implying a Fermi surface. And, this peak becomes sharper and its location of momentum is larger when we increase the Lorentz violation parameter $\eta$, as shown in the middle and right plots.  This phenomena is more obvious in figure  \ref{fig-d2z15etac} where we fix a tiny frequency $\omega=10^{-5}$ and plot the behavior of $G_{2}$ by changing the momentum.  It means that $\eta$ compensates the effect of the Lifshitz exponent which suppresses the height of peak.
From this perspective, this observation may explain why it is more reasonable to introduce Lorentz violation fermions to our black hole solution to holographically study the spectral in the boundary theory since we always expect sharp peak at $\omega=0$ to define the Fermi-like surface.
\begin{figure}[h]
  \centering
  \includegraphics[width=.4\textwidth]{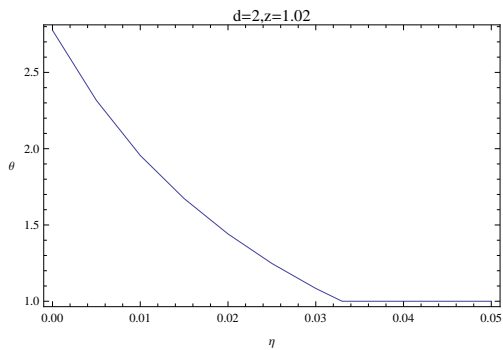}\hspace{1cm}
  \caption{The relation between $\theta$ and $\eta$.}\label{fig-d2z15etat}
\end{figure}
From figure \ref{fig-d2z15etac}, we can roughly read off the location of the peak, which denotes the Fermi momentum $k_F$ of the Fermi surface.  With more careful calculation,  we plot the related values of fermi momentum $k_F$ for different $\eta$ in figure \ref{fig-d2z15etakf}, which shows that $k_F$ increases as the Lorentz violation of Fermions is enhanced.

After we determine  the fermi momentum, we can directly calculate the dispersion relation which is defined as \cite{f3}
\begin{eqnarray} \label{eq-dispersion}
\tilde{\omega}(\tilde{k})\propto \tilde{k}^{\theta}, \quad {\rm
with} \quad \theta  = \begin{cases} \frac{1}{2 \nu(k_F)} & \mathrm{for}~~
\nu(k_F) < \frac{1}{2}\cr
            1 & \mathrm{for}~~\nu(k_F) > \frac{1}{2}
\end{cases}~.
\end{eqnarray}
where $\tilde{k}=k-k_F$ and $\tilde{\omega}$ denote the shift away form the fermi surface with $(k=k_F, \omega\to 0)$, and $\nu(k)$ is defined in equation \eqref{eq-nu}.
Substituting the determined Fermi momentum into (\ref{eq-dispersion}), we compute the exponent $\theta$ of the dispersion relation and the results are shown in figure \ref{fig-d2z15etat}. When $\eta$ is smaller than $0.033$, the exponent $\theta$ decreases with the increasing of $\eta$ and it is always higher than unit, which implies that the low energy excitation is non-Fermi Liquid. On the other side, when $\eta$ is larger than $0.033$,  the low energy excitation may behave as Fermi Liquid because $\theta$ keeps at unit. Our results imply that when the parameter of Lorentz violation of fermions reach strong enough, there may be a phase transition from non-Fermi liquid to  Fermi liquid in  boundary fermionic system. Thus, in order to holographically describe the Fermi liquid in non-relativistic HL gravity, it is more reasonable to introduce Lorentz-violation fermions.

%%%%%%%%
\section{Conclusion and Discussion}\label{sec-conclusion}
In this paper, we analytically solved the equations of motion for the coupling system between non-relativistic gauge field and HL gravitational action. We obtained a new charged Lifshitz black hole solution, which in the UV limit behaves the Lifshitz geometry \eqref{eq-LifshitzMetric} with a dynamical scaling between space and time, while at zero temperature, the near horizon geometry is $AdS_2\times\mathbb{R}^{D-2}$. Here we did not explore more detailed features of our new black hole, such as thermodynamics and  various stabilities, etc., but we will present this study somewhere else in the near future. Moreover, it would be very interesting to
study whether our bulk theory admits the hyperscaling violating black hole solution \cite{Alishahiha:2012qu,Alishahiha:2014cwa}, which has been holographically applied in condensed matter system, please see \cite{Charmousis:2010zz} and thereby in.

Then, by introducing the probed Lorentz-violation Fermions into the background, we studied the properties of the fermionic sector dual to the charged HL gravity theory.  We fixed $z=1.02$ and  focused on the effect of the Lorentz-violation feature on the Fermi momentum and the dispersion relation. We found that for stronger Lorentz-violation of the fermions, the peak in the spectral function, which is suppressed by the dynamical exponent,  is enhanced and the Fermi momentum is larger. Meanwhile, the stronger violation makes the exponent of dispersion relation lower until it reaches a critical value $\eta\simeq 0.033$, the dispersion relation is kept linear even though we further enhance the violation, which means that at the critical value $\eta\simeq 0.033$, the type of low energy excitation will transit from non-Fermi liquid to Fermi Liquid.
We also argued that in order to describe a Fermi type Liquid of the low energy excitation, it is more reasonable to add  the Lorentz-violation  fermions in the non-relativistic charged HL gravity.
It would be very interesting to further find  an interpretation for the violation parameter $\eta$ consistent with the Luttinger theorem, which states that the charge density is equal to the volume enclosed by the Fermi surface.

Note that in the fermionic sector we studied, we chose the boundary condition \eqref{eq-SR} for the Dirac field. Since all fields in our system are non-relativistic, we have strong motivation to choose the Lorentz violation boundary condition for  the bulk fermion field
\begin{eqnarray} \label{boundary2}
S_{bdy}&=&\frac{1}{2}\int_{\partial\mathcal{M}}d^3x\sqrt{-gg^{rr}}\bar{\phi}\Gamma^1\Gamma^2\phi~.
\end{eqnarray}
With the above boundary condition, it was first addressed in \cite{1108.1381} that the spectral function of the dual
holographic fermionic system presented a flat band of gapless excitation.  Then we follow the discussion of \cite{1108.1381}, in this case, the retarded Green function is a kind of recombination of  Green functions of the relativistic fixed point as
\begin{eqnarray} \label{Greenfunction2}
G_{R}=\left(
\begin{array}{cc}
\frac{2G_1G_2}{G_1+G_2} & \frac{G_1-G_2}{G_1+G_2} \\
 \frac{G_1-G_2}{G_1+G_2} & \frac{-2}{G_1+G_2} \\
\end{array}
\right)
\end{eqnarray}
which was firstly obtained in \cite{1110.4559}. Due to the symmetry \eqref{eq-symmetry},  the spectral function is
\begin{equation}\label{spectralfunction-NR}
A_{NR}(\omega,k)=\rm Im~{Tr[G_R]}=Im~ \left[\frac{4
G_I}{1-G_{I}^2}\right]~.
\end{equation}
However, we did not find Fermi surface in this case of our system. The physics behind this phenomena calls for further research, and we hope that  further study on the features of our black hole solution may give some insight.

\begin{acknowledgments}
X.M. Kuang is supported by Chilean FONDECYT grant No.3150006.
F.W. Shu is supported in part by the National Natural Science Foundation of China under Grant No.11465012, the Natural Science Foundation of Jiangxi Province under Grant No.20142BAB202007 and the 555 talent project of Jiangxi Province.
\end{acknowledgments}

%\bibliographystyle{../preprint}
%\bibliography{../Bib/QuantGra.bib}

\begin{thebibliography}{10}
%%%%%holography%%%%%%%%%%%%%%%%%%%%%%%%%
\bibitem{maldacena1} J. Maldacena,``The large-N limit of superconformal field theories and supergravity'', Adv.
Theor. Math. Phys. {\bf2} (1998) 231 [Int. J. Theor. Phys. {\bf38} (1999) 1113], [arXiv:hep-th/9711200].

\bibitem{agmoo} O. Aharony, S.~S. Gubser, J. Maldacena, H. Ooguri and Y. Oz,``Large N field theories, string
theory and gravity'', Phys. Rept. {\bf323} (2000) 183, [arXiv:hep-th/9905111].

\bibitem{gkp} S. Gubser, I.R. Klebanov and A.M. Polyakov,``Gauge theory correlators from noncritical
string theory'', Phys. Lett. B {\bf428} (1998) 105, [arXiv:hep-th/9802109].

\bibitem{witten} E. Witten,``Anti-de Sitter space and holography'', Adv. Theor. Math. Phys. {\bf2} (1998) 253,
[arXiv:hep-th/9802150] .

\bibitem{hartnoll}S. A. Hartnoll,``Lectures on holographic methods for condensed matter physics'', Class. Quant. Grav. {\bf26} (2009) 224002, [arXiv:0903.3246].

\bibitem{herzog}C. P. Herzog,``Lectures on Holographic Superfluidity and Superconductivity'', J. Phys. A {\bf42} (2009) 343001, [arXiv:0904.1975].

\bibitem{mcgreevy} J. McGreevy,``Holographic duality with a view toward many-body physics'', Adv. High Energy
Phys. {\bf2010} (2010) 723105, [arXiv:0909.0518].

\bibitem{Zaanen} J. Zaanen, Y. W. Sun, Y. Liu and K. Schalm, ``Holographic Duality in Condensed Matter Physics,"  Cambridge University Press 2015.
%%%%%%%horava%%%%%%%%%%%%%%%%%%%%%%%%%%

%\cite{Kachru:2008yh}
\bibitem{Kachru:2008yh}
  S.~Kachru, X.~Liu and M.~Mulligan,
  ``Gravity duals of Lifshitz-like fixed points'',
  Phys.\ Rev.\ D {\bf 78} (2008) 106005
  %doi:10.1103/PhysRevD.78.106005
  [arXiv:0808.1725 [hep-th]].

%\cite{Horava:2009uw,Horava:2011gd}
\bibitem{Horava:2009uw}
  P.~Horava,``Quantum Gravity at a Lifshitz Point'',
  %``Quantum Gravity at a Lifshitz Point,''
  Phys.\ Rev.\ D {\bf 79} (2009) 084008,
  [arXiv:0901.3775 [hep-th]].

%\cite{Horava:2011gd}
\bibitem{Horava:2011gd}
  P.~Horava,
  ``General Covariance in Gravity at a Lifshitz Point'',
  Class.\ Quant.\ Grav.\  {\bf 28} (2011) 114012,
  %doi:10.1088/0264-9381/28/11/114012
  [arXiv:1101.1081 [hep-th]].

\bibitem{TP}  T. Pavlopoulos, ``Breakdown of Lorentz invariance", Phys. Rev. {\bf159} (1967) 1106.

\bibitem{CN}  S. Chadha and H.B. Nielsen, ``Lorentz invariance as a low-energy phenomenon", Nucl. Phys.
B {\bf217} (1983) 125.

\bibitem{Griffin:2012qx}
  T.~Griffin, P.~Hor\v{a}va and C.~M.~Melby-Thompson,
  ``Lifshitz Gravity for Lifshitz Holography'',
  Phys.\ Rev.\ Lett.\  {\bf 110} (2013) 081602,
  %doi:10.1103/PhysRevLett.110.081602
  [arXiv:1211.4872 [hep-th]].

  %\cite{Janiszewski:2012nf,Janiszewski:2012nb}
\bibitem{Janiszewski:2012nf}
  S.~Janiszewski and A.~Karch,
  ``String Theory Embeddings of Nonrelativistic Field Theories and Their Holographic Horava Gravity Duals'',
  Phys.\ Rev.\ Lett.\  {\bf 110} (2013) 081601,
  %doi:10.1103/PhysRevLett.110.081601
  [arXiv:1211.0010 [hep-th]].

%\cite{Janiszewski:2012nb}
\bibitem{Janiszewski:2012nb}
  S.~Janiszewski and A.~Karch,
  ``Non-relativistic holography from Horava gravity'',
  JHEP {\bf02} (2013) 123 ,
  [arXiv:1211.0005 [hep-th]].

\bibitem{Alishahiha:2012iy}
  M.~Alishahiha and H.~Yavartanoo,
  ``Conformally Lifshitz solutions from Horava Lifshitz Gravity'',
  Class.\ Quant.\ Grav.\  {\bf 31} (2014) 095008,
  %doi:10.1088/0264-9381/31/9/095008
  [arXiv:1212.4190[hep-th]].

\bibitem{wang2014}
F.-W. ~Shu, K. ~Lin, A. ~Wang, Q. ~Wu,
``Lifshitz spacetimes, solitons, and generalized BTZ black holes in quantum gravity at a Lifshitz point'',
JHEP {\bf04} (2014) 056, [arXiv:1403.0946[hep-th]];
K. ~Lin, F.-W. ~Shu, A. ~Wang, Q. ~Wu,
``High-dimensional Lifshitz-type spacetimes, universal horizons, and black holes in Horava-Lifshitz gravity'',
 Phys.\ Rev.\ D {\bf91} (2015) 044003, [arXiv:1404.3413[hep-th]].

\bibitem{Lin}
K. ~Lin, E. ~Abdalla, A. ~Wang,
``Holographic superconductors in Horava-Lifshitz gravity'',
 Int.\ J.\ Mod.\ Phys.\ D {\bf24} (06) (2015) 1550038, [arXiv:1406.4721[hep-th]].

\bibitem{Lin1}
X. Wang, J. Yang, M. Tian, A. Wang, Y. Deng, G. Cleaver,``Effects of high-order operators in non-relativistic Lifshitz holography'',
Phys. Rev. D {\bf91} (2015) 064018, [arxiv:1407.1194[hep-th]]

\bibitem{Luo}
C. -J. ~Luo, X.-M. ~Kuang, F.-W. ~Shu,
``Lifshitz holographic superconductor in Horava-Lifshitz gravity'',
Phys.\ Lett.\ B {\bf759} (2016) 184, [arXiv:1605.03260[hep-th]].

%\cite{Lopes:2015bra}
\bibitem{Lopes:2015bra} 
  D.~V.~Lopes, A.~Mamiya and A.~Pinzul,
  ``Infrared Horava?Lifshitz gravity coupled to Lorentz violating matter: a spectral action approach,''
  Class.\ Quant.\ Grav.\  {\bf 33}, no. 4, 045008 (2016)
  %doi:10.1088/0264-9381/33/4/045008
  [arXiv:1508.00137 [hep-th]].
 \bibitem{SMEaction}V. A. Kostelecky, ``Gravity, Lorentz violation, and the standard model'', Phys.Rev. D {\bf69} (2004) 105009, arXiv:hep-th/0312310 [hep-th].
 %%%%%fermion-pionner%%%%%%%%%%%%%%%%%%%%%%%%%%%%%%%%%%%%
 \bibitem{f1}S. S. Lee, ``A Non-Fermi Liquid from a Charged Black Hole; A Critical Fermi Ball'',
Phys. Rev. D {\bf79} (2009) 086006,
[arXiv:0809.3402 [hep-th]].
\bibitem{f2} H. Liu, J. McGreevy and D. Vegh, ``Non-Fermi liquids from holography'',
 Phys. Rev. D {\bf83} (2011) 065029,
[arXiv:0903.2477 [hep-th]].
\bibitem{f3} T. Faulkner, H. Liu, J. McGreevy and D. Vegh, ``Emergent quantum criticality, Fermi surfaces, and AdS$_2$'', Phys. Rev. D {\bf83} (2011) 125002,
[arXiv:0907.2694 [hep-th]].
\bibitem{f4} M. Cubrovic, J. Zaanen and K. Schalm, ``String Theory, Quantum Phase Transitions and the Emergent Fermi-Liquid'', Science 325 (2009) 439,
[arXiv:0904.1993 [hep-th]].
\bibitem{IL}N. Iqbal and H. Liu, ``Real-time response in AdS/CFT with application to spinors'',
Fortsch. Phys. {\bf57} (2009) 367,
[arXiv:0903.2596 [hep-th]].

%\cite{Hartnoll:2011dm,Cubrovic:2011xm}
\bibitem{Hartnoll:2011dm}
  S.~A.~Hartnoll, D.~M.~Hofman and D.~Vegh,
  ``Stellar spectroscopy: Fermions and holographic Lifshitz criticality,''
  JHEP {\bf 1108}, 096 (2011)
  %doi:10.1007/JHEP08(2011)096
  [arXiv:1105.3197 [hep-th]].
%\cite{Cubrovic:2011xm}
\bibitem{Cubrovic:2011xm}
  M.~Cubrovic, Y.~Liu, K.~Schalm, Y.~W.~Sun and J.~Zaanen,
  ``Spectral probes of the holographic Fermi groundstate: dialing between the electron star and AdS Dirac hair,''
  Phys.\ Rev.\ D {\bf 84}, 086002 (2011)
  %doi:10.1103/PhysRevD.84.086002
  [arXiv:1106.1798 [hep-th]].

 %%%%%fermion-extension%%%%%%%%%%%%%%%%%%%%%%%%%%%%%%%
\bibitem{JPW1}J. P. Wu,  ``Holographic fermions in charged Gauss-Bonnet black hole'',  JHEP {\bf07} (2011) 106,
[arXiv:1103.3982 [hep-th]].
\bibitem{kuang1}X. M. Kuang, B. Wang, J. P. Wu, ``Dipole coupling effect of holographic fermion in the background of charged Gauss-Bonnet AdS black hole'',
JHEP {\bf07} (2012) 125, [arXiv:1205.6674[hep-th]].
\bibitem{kuang2}
X. M. Kuang, B. Wang, J. P. Wu, ``Dynamical gap from holography in the charged dilaton black hole'', Class. Quantum Grav. {\bf30} (2013) 145011, [arXiv:1210.5735 [hep-th]].
\bibitem{JPW2}J. P. Wu, ``Some properties of the holographic fermions in an extremal charged dilatonic black hole''
, Phys. Rev. D {\bf84} (2011) 064008,
[arXiv:1108.6134 [hep-th]].

%\cite{Alishahiha:2012nm}
\bibitem{1201.1764}
  M.~Alishahiha, M.~R.~Mohammadi Mozaffar and A.~Mollabashi,
  ``Fermions on Lifshitz Background,''
  Phys.\ Rev.\ D {\bf 86}, 026002 (2012)
% doi:10.1103/PhysRevD.86.026002
  [arXiv:1201.1764 [hep-th]].
\bibitem{FLQ1}
L. Q. Fang, X. H. Ge, X. M. Kuang,
``Holographic fermions in charged Lifshitz theory'', Phys. Rev. D {\bf 86} (2012) 105037, [arXiv:1201.3832 [hep-th]].

\bibitem{FLQ2}
 L. Q. Fang, X. H. Ge, X. M. Kuang,
``Holographic fermions with running chemical potential and dipole coupling'', Nucl. Phys. B {\bf877} (2013) 807, [arXiv:1304.7431[hep-th]].

\bibitem{FLQ3}
 L. Q. Fang, X. H. Ge, J. P. Wu, H. Q. Leng,
``Anisotropic Fermi surface from holography'', Phys. Rev. D {\bf 91} (2015) 126009, [arXiv:1409.6062[hep-th]].

%\cite{Fan:2013tpa,Fan:2013zqa}
\bibitem{Fan:2013tpa}
  Z.~Fan,
  ``Holographic fermions in asymptotically scaling geometries with hyperscaling violatio'',
  Phys.\ Rev.\ D {\bf 88} (2013) 026018 ,
  %doi:10.1103/PhysRevD.88.026018
  [arXiv:1303.6053 [hep-th]].
%\cite{Fan:2013zqa}
\bibitem{Fan:2013zqa}
  Z.~Fan,
  ``Dynamic Mott gap from holographic fermions in geometries with hyperscaling violation'',
  JHEP {\bf08} (2013) 119 ,
  %doi:10.1007/JHEP08(2013)119
  [arXiv:1305.1151 [hep-th]].

\bibitem{1108.1381}
  J.~N.~Laia and D.~Tong,
  ``A Holographic Flat Band'',
  JHEP {\bf11} (2011) 125,
  [arXiv:1108.1381 [hep-th]].

\bibitem{1110.4559}
  W.~J.~Li and H.~Zhang,
  ``Holographic non-relativistic fermionic fixed point and bulk dipole coupling'',
  JHEP {\bf11} (2011) 018,
  [arXiv:1110.4559 [hep-th]].

  %\cite{Li:2011sh}
\bibitem{Lin2}
A. Borzou, K. Lin, A. Wang,``Static electromagnetic fields and charged black holes in general covariant theory of Horava-Lifshitz gravity'',
JCAP {\bf02} (2012) 025, [arxiv:1110.1636[hep-th]].
\bibitem{1111.3783}
  W.~J.~Li, R.~Meyer and H.~b.~Zhang,
  ``Holographic non-relativistic fermionic fixed point by the charged dilatonic black hole,''
  JHEP {\bf01} (2012) 153
  %doi:10.1007/JHEP01(2012)153
  [arXiv:1111.3783 [hep-th]].
\bibitem{1409.2945}
  X.~M.~Kuang, E.~Papantonopoulos, B.~Wang and J.~P.~Wu,
 ``Formation of Fermi surfaces and the appearance of liquid phases in holographic theories with hyperscaling violation'',
  JHEP {\bf11} (2014) 086,
  [arXiv:1409.2945 [hep-th]].


\bibitem{1010.3238}
  M.~Edalati, R.~G.~Leigh and P.~W.~Phillips,
  ``Dynamically Generated Mott Gap from Holography'',
  Phys.\ Rev.\ Lett.\  {\bf 106} (2011) 091602,
  [arXiv:1010.3238 [hep-th]].

\bibitem{1012.3751}
  M.~Edalati, R.~G.~Leigh, K.~W.~Lo and P.~W.~Phillips,
  ``Dynamical Gap and Cuprate-like Physics from Holography'',
  Phys.\ Rev.\ D {\bf 83} (2011) 046012,
  [arXiv:1012.3751 [hep-th]].

 \bibitem{1102.3908}
  D.~Guarrera and J.~McGreevy,
  ``Holographic Fermi surfaces and bulk dipole couplings'',
  [arXiv:1102.3908 [hep-th]].

\bibitem{1411.5627}
  X.~M.~Kuang, E.~Papantonopoulos, B.~Wang and J.~P.~Wu,
 ``Dynamically generated gap from holography in the charged black brane with hyperscaling violation'',
  JHEP {\bf04} (2015) 137,
  [arXiv:1411.5627 [hep-th]].

\bibitem{1405.1041}
G. Vanacore, P. W. Phillips, ``Minding the Gap in Holographic Models of Interacting Fermions'', Phys. Rev. D {\bf 90} (2014) 044022, [arXiv:1405.1041[hep-th]].

\bibitem{1404.4010}
J. Alsup, E. Papantonopoulos, G. Siopsis, K. Yeter, ``Duality between zeroes and poles in holographic systems with massless fermions and a dipole coupling'',
Phys. Rev. D {\bf 90} (2014) 126013,[arXiv:1404.4010[hep-th]].

%\cite{Hartnoll:2012rj,Liu:2012tr}
\bibitem{Hartnoll:2012rj}
  S.~A.~Hartnoll and D.~M.~Hofman,
  ``Locally Critical Resistivities from Umklapp Scattering,''
  Phys.\ Rev.\ Lett.\  {\bf 108}, 241601 (2012)
  %doi:10.1103/PhysRevLett.108.241601
  [arXiv:1201.3917 [hep-th]].

%\cite{Liu:2012tr}
\bibitem{Liu:2012tr}
  Y.~Liu, K.~Schalm, Y.~W.~Sun and J.~Zaanen,
  ``Lattice Potentials and Fermions in Holographic non Fermi-Liquids: Hybridizing Local Quantum Criticality,''
  JHEP {\bf 1210}, 036 (2012)
  %doi:10.1007/JHEP10(2012)036
  [arXiv:1205.5227 [hep-th]].

\bibitem{1304.2128}
  Y.~Ling, C.~Niu, J.~P.~Wu, Z.~Y.~Xian and H.~b.~Zhang,
  ``Holographic Fermionic Liquid with Lattices'',
  JHEP {\bf 07} (2013) 045,
  [arXiv:1304.2128 [hep-th]].

\bibitem{1410.7323}
  Y.~Ling, P.~Liu, C.~Niu, J.~P.~Wu and Z.~Y.~Xian,
  ``Holographic fermionic system with dipole coupling on Q-lattice'',
  JHEP {\bf12} (2014) 149,
  [arXiv:1410.7323 [hep-th]].

%\cite{Fang:2015dia}
\bibitem{Fang:2015dia}
  L.~Q.~Fang, X.~M.~Kuang, B.~Wang and J.~P.~Wu,
  ``Fermionic phase transition induced by the effective impurity in holography'',
  JHEP {\bf11} (2015) 134 ,
  %doi:10.1007/JHEP11(2015)134
  [arXiv:1507.03121 [hep-th]].




%\cite{Kimpton:2013zb}
\bibitem{Kimpton:2013zb}
  I.~Kimpton and A.~Padilla,
  ``Matter in Horava-Lifshitz gravity'',
  JHEP {\bf04} (2013) 133 ,
 % doi:10.1007/JHEP04(2013)133
  [arXiv:1301.6950 [hep-th]].

%\cite{Taylor:2008tg}
\bibitem{Taylor:2008tg}
  M.~Taylor,
  ``Non-relativistic holography'',
  [arXiv:0812.0530 [hep-th]].
  %%CITATION = ARXIV:0812.0530;%%

\bibitem{dugui2} J. Tarrio, S. Vandoren, ``Black holes and black branes in Lifshitz spacetimes'', JHEP {\bf1109} (2011) 017,
[arXiv:1105.6335].

\bibitem{dugui3} W. G. Brenna, R. B. Mann, M. Park, ``Mass and Thermodynamic Volume in Lifshitz Spacetimes'', Phys. Rev. D {\bf92} (2015) 044015 ,[arXiv:1505.06331v3 [hep-th]].

\bibitem{dugui4} Z.-Y. Fan, H. Lu,``Electrically-Charged Lifshitz Spacetimes, and Hyperscaling Violations'', JHEP {\bf04} (2015) 139, [arXiv:1501.05318v1 [hep-th]].

\bibitem{dugui5} Z. Y. Fan and H. Lu,``Charged Black Holes in Colored Lifshitz Spacetimes'', Phys. Lett.
B {\bf743} (2015) 290,[arXiv:1501.01727 [hep-th]].

\bibitem{dugui6} J. Tarrio, S. Vandoren, ``Black holes and black branes in Lifshitz spacetimes'', JHEP {\bf09} (2011) 017,[arXiv:1105.6335v2 [hep-th]].

\bibitem{1401.6479}S. Janiszewski, A. Karch, B. Robinson, D. Sommer, ``Charged black holes in Hor\v{a}va gravity'', arXiv:1401.6479v1 [hep-th].

%%%%%%%%%%%%%%%%%%%%%%%%%%%%%%%%%%%%%%%%%%%%%%%%%

\bibitem{gammamatrices} T. Faulkner, G. T. Horowitz, J. McGreevy, M. M. Roberts, D. Vegh,
``Photoemission `experiments' on holographic superconductors'',
JHEP {\bf1003} (2010) 121 , [arXiv:0911.3402].

%\cite{Alishahiha:2012qu,Alishahiha:2014cwa,Charmousis:2010zz}
\bibitem{Alishahiha:2012qu}
  M.~Alishahiha, E.~O Colgain and H.~Yavartanoo,
  ``Charged Black Branes with Hyperscaling Violating Factor,''
  JHEP {\bf 1211}, 137 (2012)
  %doi:10.1007/JHEP11(2012)137
  [arXiv:1209.3946 [hep-th]].

%\cite{Alishahiha:2014cwa}
\bibitem{Alishahiha:2014cwa}
  M.~Alishahiha, A.~F.~Astaneh and M.~R.~Mohammadi Mozaffar,
  ``Thermalization in backgrounds with hyperscaling violating factor,''
  Phys.\ Rev.\ D {\bf 90}, no. 4, 046004 (2014)
  %doi:10.1103/PhysRevD.90.046004
  [arXiv:1401.2807 [hep-th]].

  %\cite{Charmousis:2010zz}
\bibitem{Charmousis:2010zz}
  C.~Charmousis, B.~Gouteraux, B.~S.~Kim, E.~Kiritsis and R.~Meyer,
  ``Effective Holographic Theories for low-temperature condensed matter systems,''
  JHEP {\bf 1011}, 151 (2010)
  %doi:10.1007/JHEP11(2010)151
  [arXiv:1005.4690 [hep-th]].
\end{thebibliography}

\end{document}